\documentclass[11pt,a4paper]{article}
\pdfoutput=1
\usepackage{jheppub}
\usepackage{wrapfig}
\usepackage{amsmath}
\usepackage{amsfonts}
\usepackage{amssymb}
\usepackage{graphicx}
\newcommand{\Tr}{\operatorname{Tr}}

\title{The Confining Transition in the Bosonic BMN Matrix Model}
\author[a]{Yuhma Asano,}
\author[b]{Samuel Kov\'a\v{c}ik,}
\author[b]{Denjoe O'Connor,}
\affiliation[a]{KEK Theory Center, High Energy Accelerator Research Organization,\\
1-1 Oho, Tsukuba, Ibaraki 305-0801, Japan.}
\affiliation[b]{School of Theoretical Physics,\\ 
       Dublin Institute for Advanced Studies, \\
       10 Burlington Road, 
       Dublin 4, Ireland.}
\emailAdd{yuhma@post.kek.jp}
\emailAdd{skovacik@stp.dias.ie}
\emailAdd{denjoe@stp.dias.ie}

\abstract{ We study the confining/deconfining phase transition in the
  mass deformed Yang-Mills matrix model which is obtained by the
  dimensional reduction of the bosonic sector of the four-dimensional
  maximally supersymmetric Yang-Mills theory compactified on the three
  sphere, i.e.~the bosonic BMN model.  The $1/D$ (with $D$ the number
  of matrices) expansion suggests that the model may have two closely
  separated transitions. However, using a second order lattice
  formulation of the model we find that for the small value of the
  mass parameter, $\mu=2$, those two apparent critical temperatures merge at
  large $N$, leaving only a single weakly first-order phase
  transition, in agreement with recent numerical results for $\mu=0$
  (the bosonic BFSS model).
  
\vspace*{.3cm}

\vspace*{.3cm}
}

\begin{document}

\maketitle

\section{Introduction}
\label{Section:Introduction}

Dimensionally reduced Yang-Mills models provide some of the simplest
candidates for understanding gauge/gravity duality and testing the
gravitational predictions for gauge theory observables. They arise in
a variety of contexts.  The supersymmetric versions such as four-dimensional
${\cal N}=4$ supersymmetric Yang-Mills theory (SYM) reduced to
one dimension, time, known as the BFSS model
\cite{deWit:1988wri,Banks:1996vh} and its maximally supersymmetric
mass deformed version known as the BMN model \cite{Berenstein:2002jq}
are conjectured to provide non-perturbative definitions of
M-theory. Both bosonic and supersymmetric models are matrix quantum
mechanical models which also arise as non-commutative deformations of
membranes and supermembranes respectively
\cite{deWit:1988wri,Kim:2006wg}. Membranes propagating on trivial
backgrounds give rise to BFSS type models \cite{deWit:1988wri} while
those on non-trivial pp-wave backgrounds \cite{Kim:2006wg} give rise
to BMN type models. They also describe the dynamics of D0-branes.

When ${\cal N}=4$ SYM is instead reduced to maximally supersymmetric
two-dimensional quantum field theory on ${\mathbb R}\times S^1$ and
then considered in a thermal bath at very high temperature the
fermions decouple and the model reduces to a purely bosonic model
which is equivalent to the bosonic BFSS model at a temperature
equivalent to the inverse period of the spatial $S^1$.  The resulting
model is the bosonic BFSS model. Alternatively considering the
dimensional reduction\footnote{In this truncation higher modes are dropped
  while preserving maximal supersymmetry.} of ${\cal N}=4$ SYM on
${\mathbb R}\times S^3$ gives rise to the BMN model.  Its bosonic
sector is studied in this paper and referred to as the bosonic BMN
model. It is also a mass deformation of the bosonic BFSS model and
corresponds to a non-commutative deformation of the bosonic membrane
propagating on a pp-wave background in eleven dimensions
\cite{Kim:2006wg}.

This family of quantum matrix models has a surprisingly rich phase
structure including confining/deconfining phase transitions
\cite{Aharony:2003sx,Furuuchi:2003sy,Semenoff:2005ei,Aharony:2005ew,Asano:2018nol}
as the temperature is varied.

When the mass parameter, $\mu$, of the model is large the model
reduces to a gauged Gaussian model, which is easily solved and has an apparent
first-order confining/deconfining phase transition
\cite{Furuuchi:2003sy,Semenoff:2005ei}. This transition would continue to
smaller $\mu$ and based on the gauge/gravity duality conjecture
for two-dimensional maximal SYM it should be connected to the
Gregory-Laflamme \cite{Gregory:1993vy,Gregory:1994bj} transition in
the dual gravitational theory at zero mass deformation.

The massless bosonic BFSS model has received much attention already in
previous studies
\cite{Aharony:2004ig,Kawahara:2007fn,Filev:2015hia,Azuma:2014cfa,Mandal:2009vz,Bergner:2019rca,Morita:2020liy}.
The earliest study was inclined to conclude that there was a single transition
\cite{Aharony:2004ig} but the $1/D$ expansion \cite{Mandal:2009vz}
suggested the system has in fact two closely separated transitions and
this was supported by numerical studies \cite{Kawahara:2007fn} at
relatively small $N$. However, more recent
studies \cite{Azuma:2014cfa, Bergner:2019rca}
at larger $N$ and new analytic results \cite{Morita:2020liy} find
evidence of a single confining/deconfining first-order phase transition.
Our study of the BMN model gives the same 
conclusion that there is only a single transition
as \cite{Aharony:2004ig,Azuma:2014cfa,Bergner:2019rca,Morita:2020liy}.

Both the bosonic BMN and BFSS models have also attracted much attention
recently \cite{Hanada:2018zxn,Bergner:2019rca} in the context of partial
deconfinement \cite{Hanada:2016pwv,Berenstein:2018lrm,Hanada:2019czd}
where it is argued that these models exhibit deconfinement of some of
the degrees of freedom but a subgroup $SU(M)\subset SU(N)$ remains
confined. This is an intriguing suggestion. However, in these matrix
models, such a phase does not seem to appear as a stable phase in the
canonical ensemble \cite{Bergner:2019rca}.
Rather what seems to happen is that the transition appears to be a
standard first-order one with a rounded critical region at finite $N$
where the system fluctuates between the two phases. In particular the
low-temperature phase has the characteristic large fluctuations at
finite $N$ of the confined phase near a Hagedorn transition. A more
detailed study is warranted in the immediate vicinity of the transition.

The principal results of the current paper are:
\begin{itemize}
\item Identification of the transition temperature with precision for $\mu=2$.
\item Verification that the model with $\mu=2$
  has a single first-order phase transition at $T_c=0.915\pm0.005$.
\item For large $N$ the system fluctuated in the transition region
  between the approximately uniform phase and the critical gapped
  phase with eigenvalue density
  $\rho(\theta)=\frac{1+\cos(\theta)}{2\pi}$.
\end{itemize}

Our results for $\mu=2$ should be close to those of $\mu=0$ where the
transition has an interpretation in terms of black strings/black hole
transitions if gauge/gravity duality holds.  With this
interpretation our results are in accord with the conclusion of
\cite{Emparan:2018bmi} where non-uniform black strings are found to be
unstable in lower dimensions and in particular for the gravity dual of
the bosonic BFSS model. The thermodynamics we find is also in accord
with other studies \cite{Dias:2017uyv,Cardona:2018shd,Ammon:2018sin}.

Furthermore support for our results comes from ${\cal N}=4$ SYM in the
large-$N$ limit on $S^3\times S^1$, where the partition function
depends only on the ratio of the radii $R$ of $S^3$ and $\beta$ of
$S^1$, i.e.~$\beta/R$, and the 't Hooft coupling. Witten argued
\cite{Witten:1998zw} that the model has a confining/deconfining
transition dual to the Hawking-Page transition in the dual
gravitational theory. If this transition occurs at sufficiently high
temperature then the fermions will decouple and one would expect that
the resulting transition would be smoothly connected to that of the
bosonic BMN and BFSS models. Our results are in accord with this picture.

The paper is organised as follows: In section \ref{Section:Model} we briefly
review the model, define our notation and list the observables we consider.
Section \ref{Section:PhaseTransitionsmu2} gives our main results
and we finish with conclusions and comments in
\ref{Section:Conclusions}.

\section{The model and observables}
\label{Section:Model}
The BMN matrix model is the quantum mechanical matrix model obtained
from a non-commutative deformation of the relativistic supersymmetric
membrane, with Nambu-Goto action in lightcone coordinates propagating
in eleven-dimensional spacetime on a pp-wave background
\cite{Berenstein:2002jq,Kim:2006wg}. The bosonic BMN model is the
corresponding bosonic model, i.e.~the BMN model without
fermions. More precisely, the model is a quantum matrix model, with
$SU(N)$ gauge symmetry, consisting of 9 Hermitian $N \times N$
matrices whose Euclidean finite temperature action is given by

\begin{align}
 S[X,A]
 =N\int \limits_0^{\beta} d \tau  \, 
 \Tr \Bigg[ &
 \frac{1}{2}D_\tau X^i D_\tau X^i 
 -\frac{1}{4} \left( [X^r,X^s]+\frac{i\mu}{3}\varepsilon^{rst}X_t \right)^2
 \label{BMNaction}
 \\
  &
 -\frac{1}{2} [X^r,X^m]^2
 -\frac{1}{4} [X^m,X^n]^2
 +\frac{1}{2}\left( \frac{\mu}{6}\right)^2X_m^2
 \nonumber
 \Bigg] ,
\end{align}
where $i=1,\cdots,9$, $r,s=1,2,3$ and $m,n=4,\cdots,9$. Also,
$\beta = 1/T$ is the inverse temperature, $\mu$ is the mass parameter
and $D_\tau\cdot=\partial_\tau\cdot-i[A,\cdot]$ is the covariant
derivative. The $SO(9)$ symmetry is explicitly broken to $SO(6)\times SO(3)$
by the mass terms and the cubic Myers term.

Since we are interested in non-perturbative results for the model,
we investigate the model numerically using the hybrid Monte Carlo algorithm described in \cite{Filev:2015hia}.
We use the lattice formulation of the model where the matrices $X^i$ are placed on lattice sites and the gauge field $A$ on links. The Euclidean time variable $\tau$ is discretised as $\tau \rightarrow \beta k / \Lambda$, where $k = 1,\cdots , \Lambda$. We use the second order discretisation of the
kinetic term discussed in \cite{Asano:2018nol, Asano:2019pre} where the
quartic term in the momentum expansion of the lattice Laplacian is set to zero.
Without loss of generality the coupling constant has been fixed to $1$ and all
dimensionful quantities are expressed in these natural units.

The lattice model depends on four parameters: $\mu$, $\beta$, $\Lambda$ and
$N$, the last two of which are to be sent to infinity to obtain the continuum,
large-$N$ limit.

Mean values of an observable $\mathcal{O}$ are defined by
integration over the ten (or more generally $D+1$)
Hermitian matrices $X^r,X^m,A$ via 
\begin{eqnarray}
\left\langle \mathcal{O} \right\rangle = \frac{\int [dX][dA] \ \mathcal{O} \ e^{-S[X,A]}}{Z}, \ Z = \int  [dX][dA] e^{-S[X,A]}.
\end{eqnarray}
In practice, the gauge field $A$ is fixed to be diagonal
and time independent with a consequent Vandermonde determinant
in the measure as discussed in \cite{Filev:2015hia}.

We measure the standard set of observables: the energy $E$, the
specific heat $C_{\rm v}$, the `extent' observable $\langle R^2\rangle $
and the Polyakov loop $\langle |P|\rangle$ which serves
as an order parameter in the confining/deconfining transition. We
also measure the Myers observable, $M$, which is crucial in the full
supersymmetric model where, at low temperatures, the fermionic terms
stabilise three of the matrices into fuzzy sphere
configurations \cite{Asano:2018nol}, but as discussed below we
find no such stable fuzzy sphere configurations for the
bosonic model.

Our principal observables are defined as

\begin{eqnarray} \nonumber
E &=& \frac{1}{N^2} (-\partial_\beta) \log Z = \frac{1}{N^2}\langle \mathcal{O}_E\rangle, \\ \nonumber
C_{\rm v} &=& \frac{\beta^2}{N^2} \partial_\beta^2 \log Z=\frac{\beta^2}{N^2}\left\langle (\mathcal{O}_E-\langle\mathcal{O}_E\rangle)^2 - \mathcal{O}'_E \right\rangle, \\
\label{obs}
\langle |P|\rangle  &=& \left\langle \frac{1}{N}\left| \Tr \left( \exp \left( i \beta A \right) \right)\right| \right\rangle , \\ \nonumber
\langle R^2\rangle &=& \left\langle \frac{1}{N \beta}  \int \limits_0^\beta d \tau \Tr\left( X^i X^i \right) \right\rangle , \\ \nonumber
M &=&  \left\langle \frac{i}{3 N \beta}  \int \limits_0^\beta d \tau \varepsilon^{rst} \Tr\left( X^r X^s X^t \right) \right\rangle ,
\end{eqnarray}
where
\begin{equation}
  \mathcal{O}_E = \frac{N}{\beta} \int \limits_0^\beta d \tau \mbox{ Tr } \left( - \frac{3}{4} [X^i,X^j]^2  - \frac{5}{6} i \varepsilon_{rst}X^r X^s X^t
  + 2 \left(\frac{\mu }{3}\right)^2 X^rX^r + 2\left(\frac{\mu }{6}\right)^2 X^mX^m\right) ,
\end{equation}
and
\begin{equation}
  \mathcal{O'}_E = \frac{N}{\beta^2} \int \limits_0^\beta d \tau \mbox{ Tr }
  \left(-\frac{3}{2} [X^i,X^j]^2  - \frac{5}{4} i \varepsilon_{rst}X^r X^s X^t
  + 2 \left(\frac{\mu }{3}\right)^2 X^rX^r + 2\left(\frac{\mu }{6}\right)^2 X^mX^m\right) .
\end{equation}
Note: As can be seen from the presence of $\mathcal{O'}_E$ in the path
integral version of the specific heat $C_{\rm v}$, the propability
distribution of $E$ as measured
in the path integral does not directly give the probability distribution of the
quantum mechanical energy. 

There are two additional observables that increase the precision of
critical temperature estimates, they will be introduced shortly.

When the mass parameter $\mu$ is very large, the model reduces to a solvable
(gauged Gaussian) model. A straightforward calculation \cite{Furuuchi:2003sy,Hadizadeh:2004bf,Semenoff:2005ei}
shows that, in the large-$N$ limit, there is a single phase
transition with critical temperature
$T_c = \frac{\mu }{6\log\left(3+2\sqrt{3}\right)}$ where
$\langle|P|\rangle$ jumps from
$\langle|P|\rangle = 0$ to $\langle|P|\rangle=1/2$ and then gradually
increases to $\langle|P|\rangle=1$ as the temperature is further increased.

For the case with $\mu = 0$, known as the bosonic BFSS model, there
are already several studies in the literature. These include a
perturbative expansion in $1/D$
\cite{Mandal:2009vz,Filev:2015hia,Takeuchi:2017wii},
where $D$ is the number of matrices ($D=9$ in our case), and numerical studies
\cite{Kawahara:2007fn,Azuma:2014cfa,Filev:2015hia,Bergner:2019rca}.
Both the $1/D$ expansion \cite{Mandal:2009vz} and earlier studies
\cite{Kawahara:2007fn} reported two closely separated critical
temperatures. The $1/D$ expansion predicts\footnote{The expression
(4.30) of \cite{Mandal:2009vz} leads to $T_{c1}=1/\beta_{c1}(9)=0.895$;
however, inverting $\beta_{c1}(D)$ and expanding it in $1/D$
yields $T_{c1}=0.891$. The same goes for the second critical temperature,
obtained as $T_{c2}=1/\beta_{c2}(9)=0.911$ from
$\beta_{c2}(D)=\beta_{c1}(D)-\frac{\ln D}{6D^{4/3}}$ with
the error being $0.002$\,.}
$T_{c1}=0.895\pm0.004$ and $T_{c2}=0.911\pm 0.002$ with
the difference in critical temperatures decreasing with
increasing $D$ as $\frac{1}{6D^{2/3}\ln D}$.  Between the two critical
temperatures, the $1/D$ expansion predicts that $\langle |P|\rangle$
should gradually increase from $0$ to $1/2$ with increasing temperature.
However, it is conceivable that
further increasing the order in the perturbative loop expansion
will close the gap resulting in its disappearance in a non-perturbative
calculation and hence show that the model has in fact a single transition.

Early numerical studies found reasonable agreement with the one-loop
$1/D$ predictions; however, a refined recent study
\cite{Bergner:2019rca} finds only one transition. They
\cite{Bergner:2019rca} find the single transition occurs between the
predicted transitions of the $1/D$ expansion and when we perform an
extrapolation using their figure 7 we estimate $T_c=0.89\pm 0.01$.  As
we will see below, our analysis of the mass deformed model with
$\mu=2$ will agree with the conclusion that there is only one
transition for $D=9$.

\begin{figure}[t] 
  \includegraphics[width=6in]{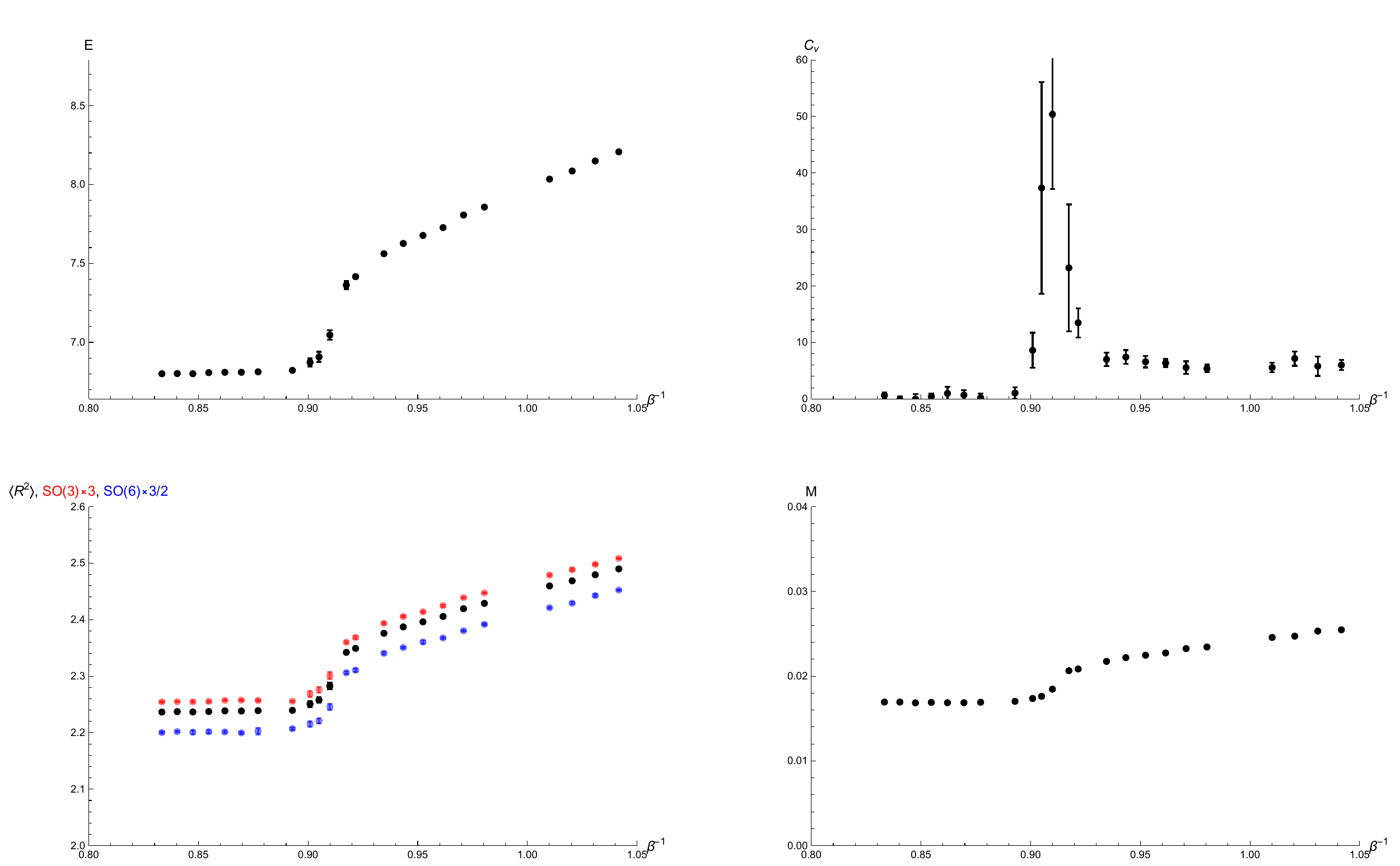}
\caption{\small The energy $E$, specific heat $C_{\rm v}$, $\langle
  R^2\rangle$ with its rescaled $SO(3)$ and $SO(6)$ components and
  the Myers observables of the model for
  $\mu = 2, N = 32\mbox{ and }\Lambda = 24$ are shown.
  The Myers observable copies the shape of $\langle
  R^2\rangle$ (and its $SO(3)$ and $SO(6)$ components) but has
  minuscule expectation values in comparison. All observables point to
  either a single or multiple transitions around $T \approx 0.91$. }
\label{Fig:Observables}
\end{figure}

The $1/D$ expansion can be easily extended to the mass deformed, bosonic BMN model, in a double expansion in $1/D$ and perturbation theory in the
cubic Myers term. Perturbation in the Myers term can be justified by its
small value shown in the Figure \ref{Fig:Observables}. As argued in \cite{Bergner:2019rca}, it appears that  $D = 9$ is not sufficiently large
to trust the $1/D$ expansion in predicting the phase transition
structure though it gives a reliable indication of the critical region.
Our analysis supports this conclusion also for non-zero $\mu$.

\begin{figure}[t] 
  \includegraphics[width=6in]{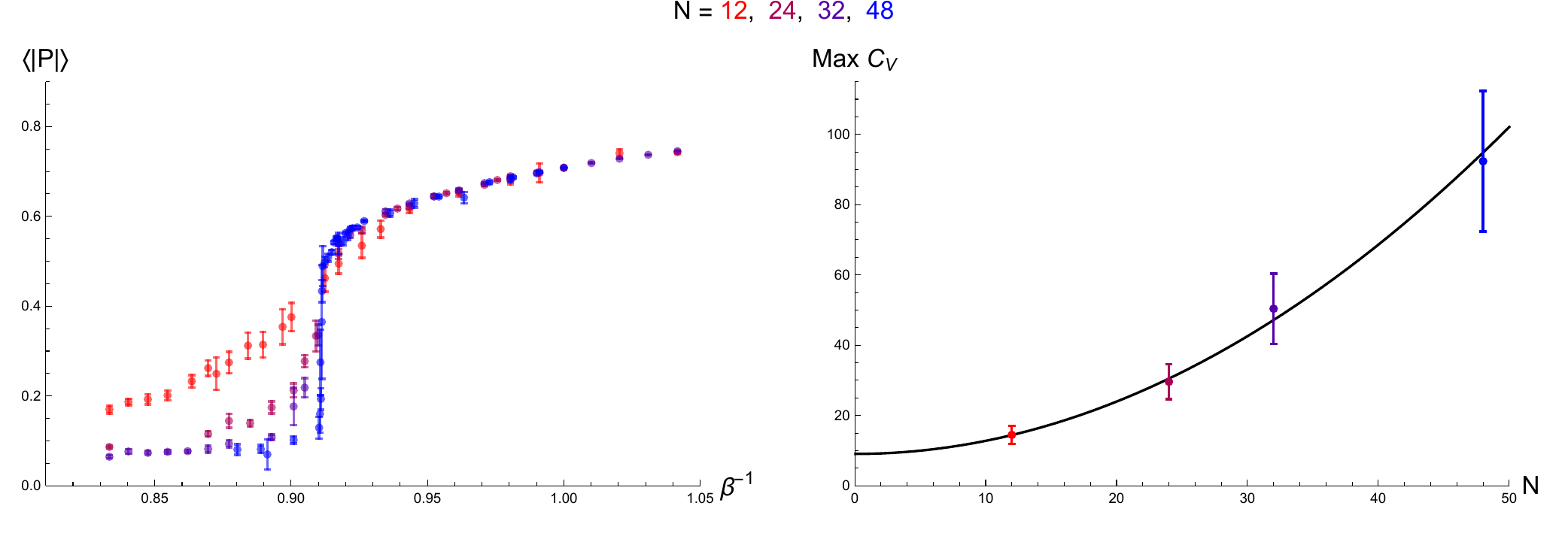}
\caption{\small The evolution of the Polyakov loop $\langle|P|\rangle$
  for $\mu = 2, \Lambda = 24$ with increasing $N$. The transition
  region becomes sharper with larger $N$ but also, due to more
  demanding simulations, the statistical errors grow. The right figure
  shows the maximum specific heat against $N$ with fit
  $C_{\rm v}^{\rm  Max}=9.1(8)+0.037(2)N^2$, adding a linear term increases errors
  and  does not improve the fit.}
\label{Fig:pols}
\end{figure}

\section{Phase transition(s) of the model with $\mu=2$}
\label{Section:PhaseTransitionsmu2}

In this paper we restrict our study to a single value of
$\mu$.  We chose $\mu = 2$ as the most interesting value, since it is
neither large (the asymptotically large mass region of the phase diagram
can be accessed analytically), nor small, being of order $1$
in natural units for the model.

Figure \ref{Fig:Observables} shows the temperature dependence of the
internal energy, $E$, its specific heat $C_{\rm v}$, $\langle R^2\rangle$
and the Myers term for $\mu = 2$, $ \Lambda = 24$ and
$N=32$ (all with jackknife error bars).  Figure \ref{Fig:pols} shows
the temperature dependence and evolution of the Polyakov loop
$\langle|P|\rangle$ with increasing $N$. The right panel of Figure
\ref{Fig:pols} shows that the peak of the specific heat grows
quadratically\footnote{A similar quadratic growth was observed for the
  Polyakov loop susceptibility in \cite{Azuma:2014cfa}.}  with $N$.
Due to the rapid increase in the number of degrees of freedom with $N$,
the simulations become more demanding and it becomes
increasingly difficult to get sufficient data to keep the errors down,
especially in the critical region.  From the figure it is clear that
the system undergoes one or more phase transitions in the vicinity of
$T_c \sim 0.91$, yet in Figure \ref{Fig:pols} it is difficult to
resolve two close transitions as expected from the $1/D$ expansion. To
do so, we analyse the transition(s) in more detail below.

In our choice of gauge fixing \cite{Filev:2015hia}, where the gauge
field is diagonal and placed on the final link,
i.e.~between $\Lambda$ and $1$,
the gauge field is fully described by a vector of angles
$-\pi < \theta_i \le \pi$ for $i=1, \dots ,N $. These are
described by a probability
distribution $\rho_N(\theta) = \frac{1}{N} \sum \limits_{i=1}^N \langle \delta (\theta -\theta_i)\rangle$, which in the large-$N$ limit gives
distribution $\rho(\theta)$.

True phase transitions occur only in the large-$N$ limit and they are
rounded at finite $N$. In the bosonic BMN model there are three
distinguishable apparent phases for finite values of $N$.
When the temperatures are very low, the distribution is approximately
uniform (up to finite-$N$ corrections). As the temperature is increased,
the distribution becomes more non-uniform and then it develops a gap. 
We will sometimes abuse terminology by referring to such transition
temperatures as critical temperatures though more strictly they are pseudo,
apparent or effective critical temperatures. This should not cause
confusion as in the end we will only identify one true
critical temperature.

Since the distribution is restricted to the periodic interval
$(-\pi,\pi]$ it is convenient to use the Fourier transformation so that
  $u_n$ is defined as the non-trivial $n$-th moment of
  $\rho(\theta)$, 
\begin{equation}
u_n = \int \limits _{-\pi}^\pi\rho(\theta) e^{i n \theta} d \theta\, .
\label{u_n}
\end{equation}
Therefore, we also define the following analogous generalizations of the
Polyakov loop as\footnote{The $U(1)$ transformation of the model
  where $A\rightarrow A+\alpha{\bf1}$ is used to remove the phase and
  one is left with the modulus.}
\begin{equation}
\langle\vert P_n\vert\rangle = \left\langle \frac{1}{N}\left| \Tr \left( \exp \left( i n \beta A \right) \right)\right| \right\rangle .
\label{P_n}
\end{equation}

Notice that the first moment $u_1$ is actually equal to the
expectation value of the Polyakov loop i.e.~$u_1=\langle\vert P\vert\rangle$.
The effective potential for
$u_1$ can be obtained, at least approximately, from the $1/D$
expansion \cite{Mandal:2009vz}. It can be expanded as a polynomial
with $D$ and temperature dependent coefficients.  It
is minimised by $u_1=0$ for $T<T_{c1}$, then grows until it reaches
 $u_1 = 1/2$.  The prediction of the $1/D$
analysis is therefore that, as the temperature is increased, beyond the
first critical temperature, $T_{c1}$, the first moment
(Polyakov loop) develops a positive expectation value.  As the
temperature is further increased  $\langle|P|\rangle$ grows reaching
$\langle|P|\rangle=1/2$ at the second critical temperature $T_{c2}$. Above this
second critical temperature the distribution becomes gapped and
the second and higher moments become non-zero.

Note that the effective potential in the large-$\mu$ limit is a quadratic
function and as the quadratic coefficient flips sign at $T_{c1}$,
the first moment $u_1$ jumps immediately to $1/2$.
Therefore, $T_{c1}= T_{c2}$ in this limit and there is a
single critical temperature. 

For generic $\mu$ at one loop the $1/D$ expansion predicts two
transitions; the first occurring when the Polyakov loop departs from
zero and the second when it reaches $1/2$, where a Gross-Witten type
transition occurs. However, since these transitions are so close and
the $\mu=0$ study indicates there is only one transition, it is
possible that there are two transitions for non-zero $\mu$ which
merge into a single transition at $\mu=0$. Alternatively, there may be
a single transition for a range or possibly all values of $\mu$.
In an effort to resolve this issue we resort to a non-perturbative
lattice study of the model.

\begin{figure}[t] 
\includegraphics[width=6in]{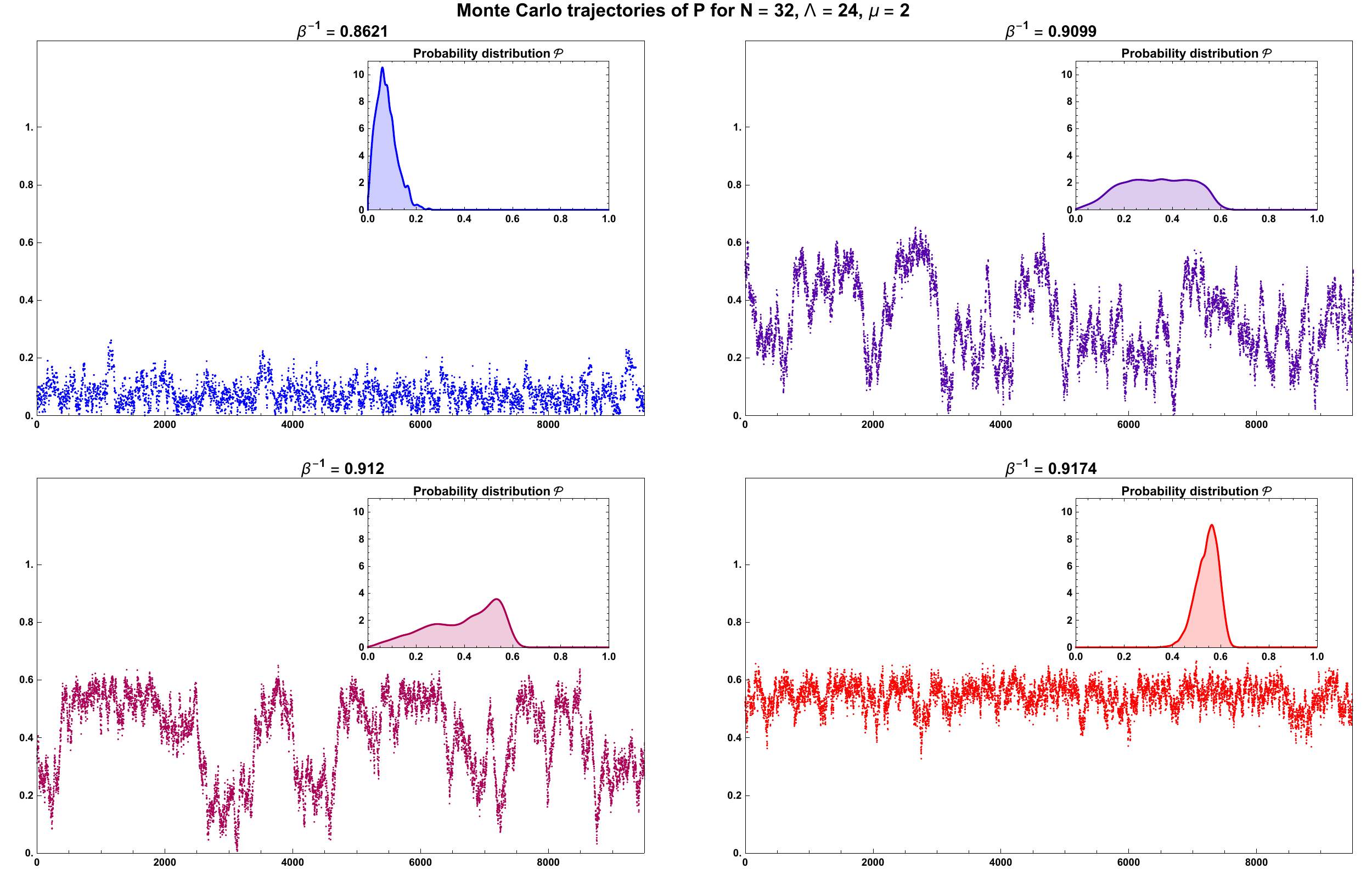}
\caption{\small Monte Carlo trajectories of the Polyakov loop,
  $\langle\vert P\vert\rangle$, for $\mu = 2, N = 32\mbox{ and } \Lambda = 24$. There
  is one clear top level and one less visible bottom level. As the
  temperature is increased, the system tends to spend more Monte Carlo
  steps in the upper one. The colours are matched with the coloured
  points in Figure \ref{Fig:Q}. We can observe that the transition is
  weakly first-order.}
\label{Fig:poltraj}
\end{figure}

To measure the first critical temperature, we need to understand the
behaviour around this transition. Monte Carlo trajectories of
$\langle\vert P\vert\rangle$ are shown in Figure
\ref{Fig:poltraj}. From the figure it seems plausible that there are
two distinct levels, suggesting a first-order phase transition. 
This behaviour becomes more articulate at higher $N$. There seems to
be one clear level around $\langle\vert P\vert\rangle = \frac{1}{2}$
and one significantly below it. Therefore, we define a new observable,
$\mathbb{P}$, defined via
\begin{equation}
  \mathbb{P}=\mathbb{P}_{\frac{1}{2}}\, ,\quad\hbox{with}\quad
  \mathbb{P}_x = \int \limits_{x}^1 \mathcal{P} (q) dq\, ,
\end{equation}
where $\mathcal{P}(q)$ is the probability distribution for the Polyakov loop.

\begin{figure}[t] 
  \includegraphics[width=6in]{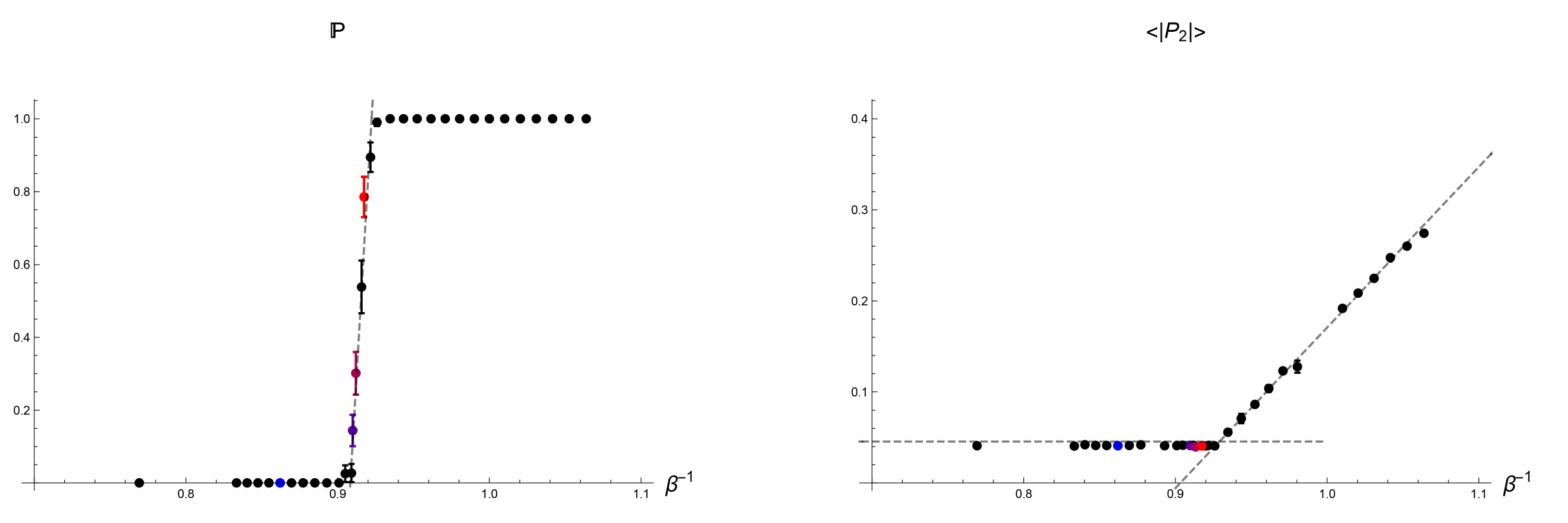}
  \caption{\small Values of $\mathbb{P}$ and $\langle\vert P_2\vert\rangle$
    for $\mu = 2, N =
  32\mbox{ and } \Lambda = 24$ and increasing value of temperature $T
  = \beta^{-1}$. The points in the transition region in the left plot
  were fit by a linear function whose slope increases with
  increasing $N$. The four coloured points are colour-matched with
  the trajectories in Figure \ref{Fig:poltraj}.}
\label{Fig:Q}
\end{figure}

Well below the transition $\mathbb{P}$ is zero, while well above it is
one and it grows very quickly around the critical
temperature,
as seen in the left panel of Figure \ref{Fig:Q}.
It is closely related to $\langle\vert P\vert\rangle$ but seems to be less
prone to finite-$N$ effects. 
The value of $x$ in $\mathbb{P}_x$ should be chosen between the value of $\langle\vert P\vert\rangle$ in the low-temperature phase ($\sim O(1/N)$) and the high-temperature phase ($\approx 0.5$);
in fact, we observed that ${\mathbb P}_{0.4}$ and
${\mathbb P}_{0.3}$ gave consistent results
to ${\mathbb P}$.
We fit the transition region of this
curve for $\mathbb{P}$ with a linear function and find its intercept with the
temperature axis to obtain $T_{c1}$.

As discussed above, the second moment, $\langle\vert P_2\vert\rangle$,
provides a strong marker for a gapped-to-ungapped
transition\footnote{$\langle\vert P_2\vert\rangle$ and higher moments
  were also discussed in \cite{Aharony:2004ig} and
  \cite{Azuma:2014cfa}.}. Higher moments would provide similar
information but with larger errors.  We present $\langle\vert P_2\vert\rangle$
versus temperature for $\mu = 2, N = 32\mbox{ and } \Lambda = 24$
in the right panel of Figure \ref{Fig:Q}.

\begin{figure}[t] 
\includegraphics[width=6in]{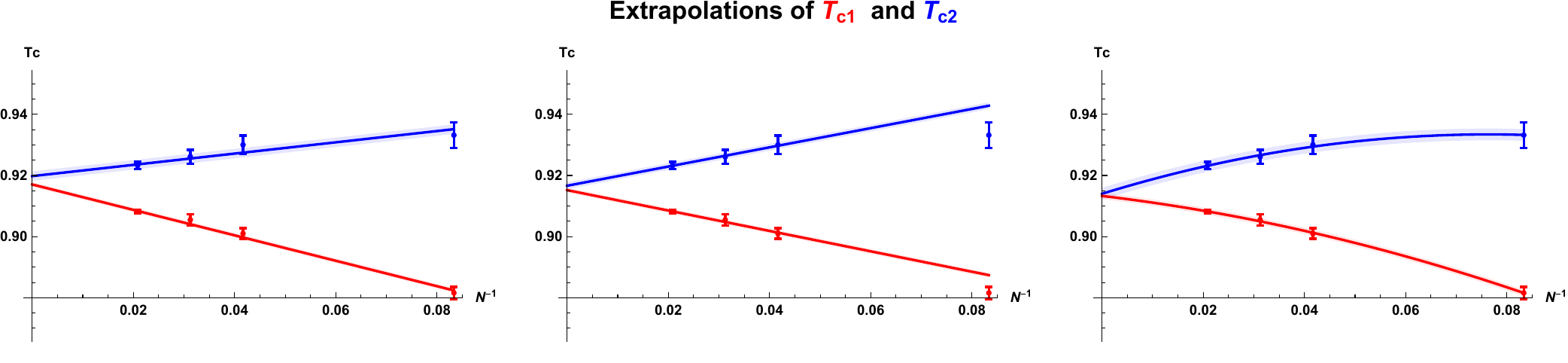}
\caption{\small Extrapolations of the critical temperatures from
  results obtained from $N = 12, 24, 32, 48$ with $\Lambda=24$.
  The left, middle and right plots use a linear fit with all data,
  a linear fit but with the $N=12$ data omitted, and a quadratic fit,
  respectively. All fits are functions of $N^{-1}$.}
\label{Fig:extrap}
\end{figure}

To identify the effective critical temperature from
$\langle\vert P_2\vert\rangle$ we fit the data above and below the
lower effective critical temperature with two linear functions.
All of the data in the low-temperature phase is well fit
by a line, and approximately twenty of the data points nearest to the transition
are fit by a line in the higher temperature region. The intercept
of these two linear fits is then taken as the effective critical temperature
(the error is the propagated error of the fitting functions).
We repeated this study with $N = 12, 24, 32, 48$ and $\Lambda=24$.
Our extrapolation of the results to large $N$ is shown in
Figure \ref{Fig:extrap}. The results depend only slightly
on the form of extrapolation function used as shown in the table:
\begin{center}
\begin{tabular}{|c|c|c|c|}
\hline 
 & Linear & Linear without $N=12$ & Quadratic \\ 
\hline 
$T_{c1}$ &$ 0.9172 (7)$ & $0.9154 (7)$ & $0.9137 (9)$ \\ 
\hline 
$T_{c2}$ & $0.919 (2)$ & $0.917 (1)$ & $  0.914 (2) $ \\ 
\hline 
\end{tabular}
\end{center}
We conclude that, in the large-$N$ limit, there is a single transition
as in the Gaussian model for large $\mu$.

\begin{wrapfigure}{l}{0.6\textwidth}
\includegraphics[width=0.59\textwidth]{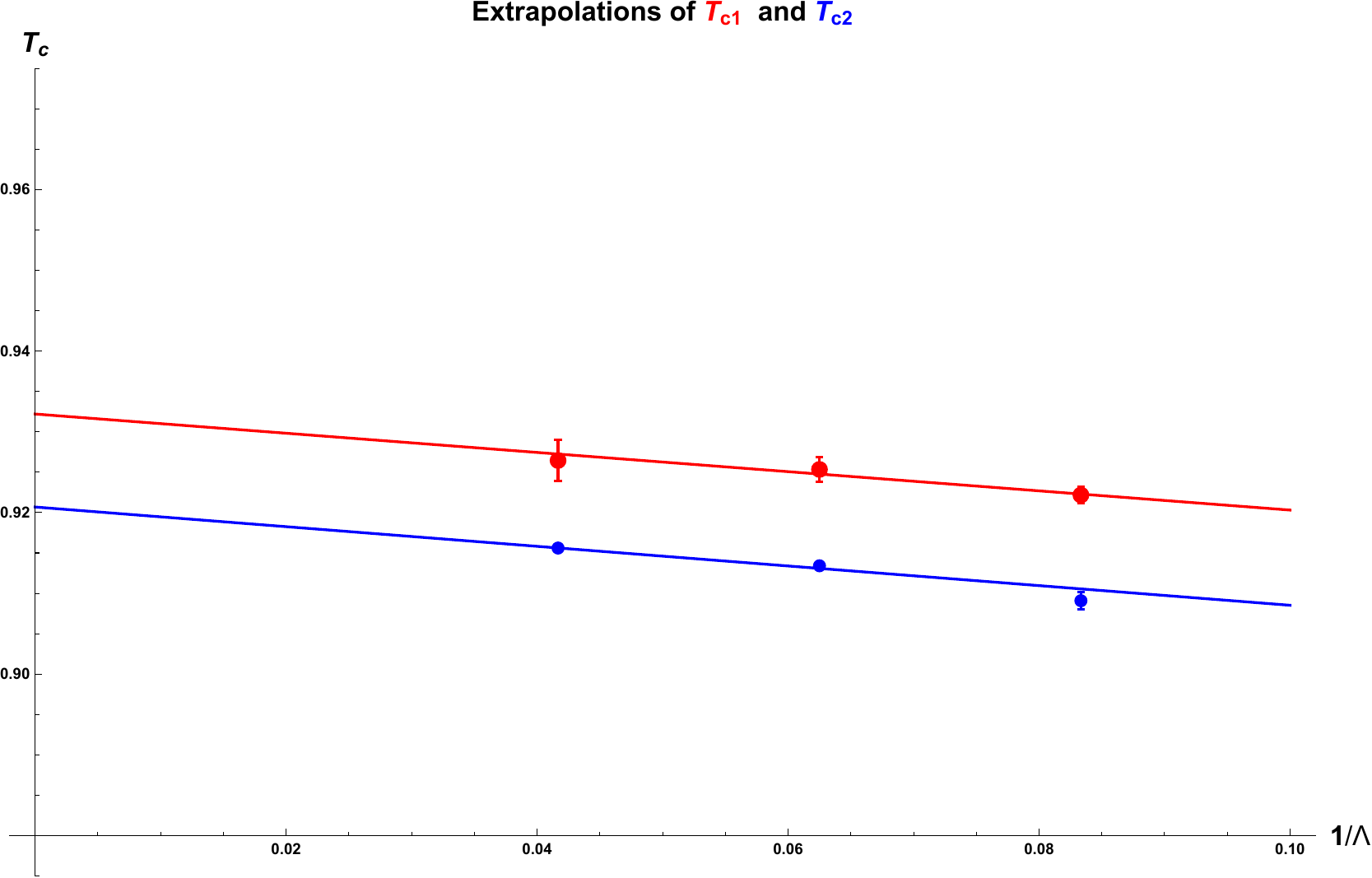}
\caption{\small Lattice dependence for $N=32$.}
\label{Fig:LambdaExtrap}
\vspace{-1.5\baselineskip}
\end{wrapfigure}

For $N=32$ extrapolation of the effective critical temperature
to the continuum limit was performed using $\Lambda=12, 16, 24$. The
measured critical temperatures are well fit in Figure \ref{Fig:LambdaExtrap}
with
\begin{eqnarray}
  T_{c1}(\Lambda) = T_{c1} (\infty) - \frac{0.12(2)}{\Lambda}\;\qquad
  \hbox{and}\qquad 
  T_{c2}(\Lambda) = T_{c2} (\infty) - \frac{0.12(3)}{\Lambda}\;.
\end{eqnarray}
The gap between the effective critical temperatures was found to scale as 
\begin{equation}
  (T_{c2} - T_{c1})(\Lambda) =  (T_{c2}-T_{c1})(\infty) + \frac{0.003(24)}{\Lambda}.
\end{equation}
For $\Lambda=24$ we see that
$(T_{c2} - T_{c1})(24) - (T_{c2} - T_{c1})(\infty) = 0.0001(10)$ and
therefore using $\Lambda=24$
seems sufficient so that lattice effects are within the errors of
the simulation.

Our conclusion is therefore that in the large-$N$ and continuum limit,
for $\mu=2$, the gap between the effective finite-$N$ critical
temperatures vanishes and there is a single, uniform-to-gapped,
first-order phase transition. From the table above and including
lattice errors we estimate that the true critical temperature of the
continuum, large-$N$, mass deformed model with $\mu=2$ is
$T_c=0.915\pm0.005$.

\bigskip


\begin{figure}[t] 
 \begin{tabular}{@{}c@{}}
\includegraphics[width=6in]{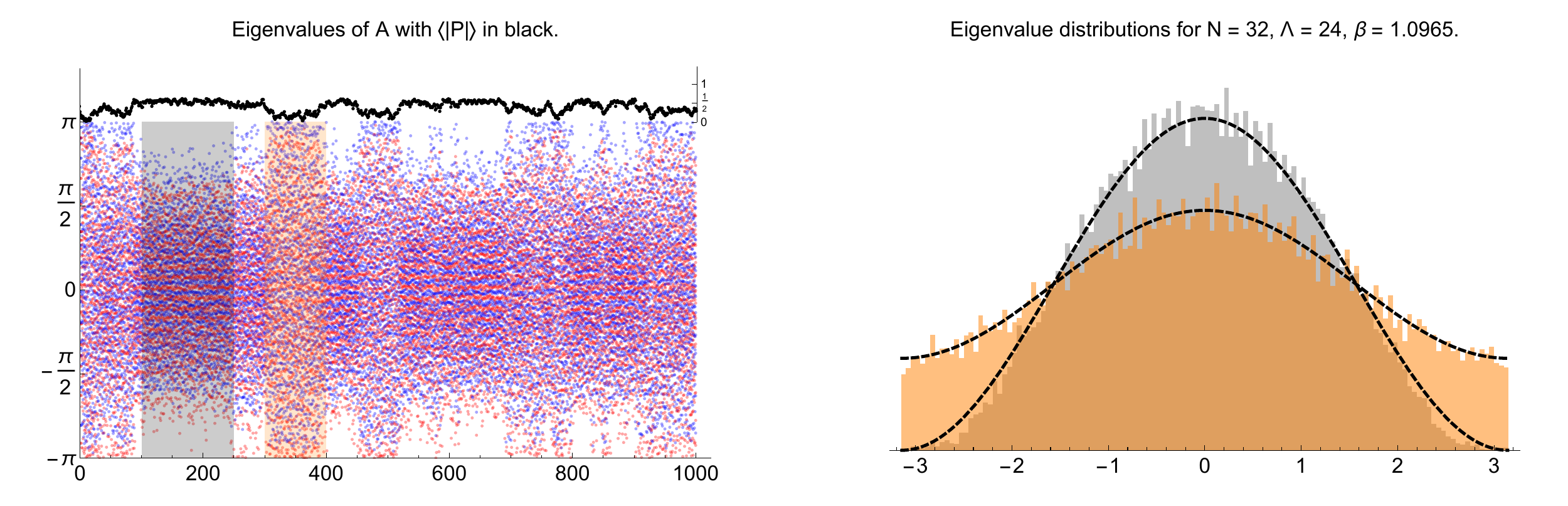}
    \end{tabular}
 \caption{\small The left figure shows the Monte Carlo evolution
   (every 10th step is shown) of the eigenvalues together with the
   evolution of the Polyakov loop $\langle\vert P\vert\rangle$.  On the right we
   plot eigenvalue distributions corresponding to eigenvalues on the
   left. The grey plot shows the distribution evaluated from
   configurations between MC times 1000 and 2500 corresponding to the
   Polyakov loop in the upper level around 0.5 while the orange plot
   shows the distribution between MC times 3000 and 4000 corresponding
   to the Polyakov loop in the lower level.
   The system is fluctuating between the approximately uniform distribution
   $\rho(\theta)=\frac{1+2\langle\vert P\vert\rangle \cos(\theta)}{2\pi}$
   with $\langle\vert P\vert\rangle=0.22$ and the critical distribution
   $\rho(\theta)=\frac{1+\cos(\theta)}{2\pi}$, represented
   by the dashed curves for the orange and grey histograms in the figure,
   respectively.  For the orange data segment
   $\langle\vert P\vert\rangle=0.22\pm0.03$ and for the grey data
   $\langle\vert P\vert\rangle=0.53\pm0.01$.}
\label{Fig:hiplotAndEVdists}
\end{figure}

\section{Conclusions}
\label{Section:Conclusions}
In this paper we find that the bosonic BMN model at $\mu=2$ undergoes
a single confining/deconfining phase transition in the large-$N$
limit. Combining this result with the result at asymptotically large
$\mu$ and the $\mu=0$ study in \cite{Bergner:2019rca} we conclude that
the most probable scenario is that there is a single first-order
transition with a $\mu$ dependent critical temperature for any value
of the mass parameter.

In contrast to this, the supersymmetric version of the model studied
in \cite{Asano:2018nol} has a rich phase structure\footnote{Its supergravity dual was studied in \cite{Costa:2014wya}} and in addition to
the confining/deconfining transition (it has not been established
whether there is more than one closely separated transitions for small
$\mu$), there also is a Myers phase transition, where the Myers
observable, \eqref{obs}, grows rapidly as the model develops fuzzy
sphere backgrounds. No such transition occurs in the bosonic case as
seen from the Myers term in Figure \ref{Fig:Observables} and the fact
that both the $SO(3)$ and $SO(6)$ sums of $\langle X^2_i\rangle$,
contributing to $\langle R^2\rangle$, behave similarly. The absence of
fuzzy spheres is not surprising since the $SO(3)$ sector of the
potential in the bosonic model (\ref{BMNaction}) is a complete square
and hence bounded below by zero\footnote{In the matrix model of this sector studied in \cite{DelgadilloBlando:2007vx,DelgadilloBlando:2008vi,DelgadilloBlando:2012xg} the transition only occurs for sufficiently negative quadratic term.} and the zero point fluctuations add an
effective positive quadratic contribution which makes the trivial
configuration the only stable vacuum.  In the supersymmetric model the
fermions cancel the bosonic zero point energy. It is therefore the
fermions that drive the supersymmetric model into a fuzzy sphere phase
and they are required to support the fuzzy sphere background.



We also find in our study, that the eigenvalue distribution for the
gauge field in the temperature range $T_{c1} < T < T_{c2}$ (where
$T_{c1}$ and $T_{c2}$ are the pseudo-critical temperatures measured at
finite $N$) can be fit by the proposed function
\begin{equation}\rho = \frac{1}{2\pi}
  \left( 1 + p \cos \theta \right), \quad 0< p < 1\, .
\label{mixeddensity}
\end{equation}
In this case the prediction for the Polyakov loop, or $u_1$ of
(\ref{u_n}), in this temperature interval is $u_1=\frac{p}{2}$.

This distribution \eqref{mixeddensity} does not simply mean the system
is in the ungapped, non-uniform phase, which is conjectured to be a
partially deconfined phase.  The distribution can be realised in
simulations via two ways: One way is that the eigenvalue distribution is
in this form at each Monte Carlo step, and the other is that this form
is realised only as an average of different kinds of distributions over
Monte Carlo time.  We observe, in this matrix model, the distribution
near the transition temperature is realised in the latter way.  By
examining the eigenvalues in detail we find that what happens is that
the eigenvalues fluctuate between the gapped and ungapped phase (see
Figure \ref{Fig:hiplotAndEVdists} and similar behaviour for $\mu=0$ in
\cite{Bergner:2019rca}) and that it is the spacing of the eigenvalues
and hence their distribution that changes.  Since Monte Carlo
simulations should realise possible physical states of the system,
superposition of the two phases is how the distribution
\eqref{mixeddensity} is realised.  Therefore it is understood as the
interpolation $\rho=(1-p)\rho_u+p \rho_c$ between the uniform
distribution $\rho_u=\frac{1}{2\pi}$ and the critical density
$\rho_c=\frac{1+\cos(\theta)}{2\pi}$, where $p$ is the probability of
finding the system in the phase with critical density $\rho_c$, which
corresponds to the endpoint density of the deconfined phase.  Although
configurations with $0<p<1/2$ in the Monte Carlo trajectories, they make
a negligible contribution to the distribution.

Note here that, although physical states are realised in simulations,
measurement of an observable does not necessarily reproduce the real
probability distribution at finite $N$.  For example, the probability
distribution of the energy observable in \eqref{obs} or any other simple
observables of energy
does not give the quantum mechanical distribution of energy.  As noted
in section \ref{Section:Model}, one can see from the specific heat
expression in (\ref{obs}) that they can be quite different.

Our study suggests that the confining/deconfining transition is a
relatively standard weakly first-order one.  As expected in such a
transition, at finite but large enough $N$, we observe that the peak of
the specific heat grows as $N^2$ (the number of degrees of freedom)
shown in Figure \ref{Fig:pols} and that the two-level features are
clearly visible\footnote{ We also observed that the two-level feature of
the transition becomes clearly visible only above $N \approx 30$, though
the growth in the peak of the specific heat in Figure \ref{Fig:pols} is
apparent for smaller $N$.  } in Figure \ref{Fig:poltraj}.  That the
transition is first-order is also in accord with the findings from the
bosonic BFSS model \cite{Azuma:2014cfa,Bergner:2019rca}.

The behaviour of the specific heat in the standard first-order
transition is well understood in statistical physics.  The
renormalization-group approach \cite{Fisher:1982xt} and other work
\cite{Landau:1984,Challa:1986sk} predict that the transition is smeared
over a region whose width goes to zero as the inverse of the number of
degrees of freedom and the specific heat has a peak with maximum which
grows in proportion to the number of degrees of freedom. This peak for
the finite system replaces the discontinuity of the specific heat in the
thermodynamic limit. The effect is easily modeled by a partition
function built from two extensive Gaussian distributions
\cite{Glaser:2017sbe} --- one around each peak.

As discussed above, the two-level features are interpreted as a mixture
of the two phases in the transition region.  The growth of the Polyakov
loop, between the two transitions, is related to the relative amount of
time the simulation spends in the upper and lower levels seen in Figure
\ref{Fig:poltraj}. The parameter $p$ in (\ref{mixeddensity}) is then
measured by the fraction of Monte Carlo time the system spends in the
level $|P|\sim 0.5$ and is closely related to our observable ${\mathbb
P}$. In the transition region ${\mathbb P}$ is surprisingly linear as
seen in Figure \ref{Fig:Q}. That the slope increases with increasing $N$
is inferred from the convergence of the two pseudo-critical temperatures
shown in Figure \ref{Fig:extrap}. The linear behaviour of ${\mathbb P}$
in the transition region is probably due to the exceptionally small
temperature range over which the finite-$N$ transition occurs.

These finite-$N$ behaviours, such as the interpolation of two kinds of
distributions in \eqref{mixeddensity} and the $N$-dependence of the
specific heat, can be understood by the double Gaussian approximation,
where the probability distribution of the Polyakov loop, an order
parameter, is described by the summation of two Gaussian distributions
centred at two values of the order parameter
\cite{Landau:1984,Challa:1986sk}.  This picture emerges as follows.  One
can transform the path-integration for the expectation value of an
observable to an integration over the order parameter by inserting
$1=\int du_1\, \delta(u_1-\frac{1}{N}\Tr\exp[i\beta A])$ and integrating
out all the matrix fields. The resultant expectation value of an
observable, $\mathcal{O}$, is written as
\begin{equation}
 \frac{1}{z}\int du_1\, \mathcal{\bar O}(u_1)e^{-f(\beta;u_1)},
\end{equation}
where $z=\int du_1\, e^{-f(\beta;u_1)}$ and
\begin{equation*}
 \mathcal{\bar O}(u_1)=e^{f(\beta;u_1)}\int [dX][dA]\,
 \mathcal{O}\,\delta(u_1-\tfrac{1}{N}\Tr\exp[i\beta A])\, e^{-S[\beta;X,A]}.
\end{equation*}
Its exact computation has not been achieved yet even for the
non-interacting gauged Gaussian model.  However, there is speculation
based on approximations by the Landau-Ginzburg model and gauge/gravity
duality
\cite{Aharony:2003sx,Aharony:2004ig,Hadizadeh:2004bf,Morita:2020liy}:
Within $0\le u_1\le 1$, $f(\beta; u_1)$ is expected to have one minimum
around $u_1=0$ at low enough temperatures and around $u_1\ge 1/2$ at
high enough temperatures but with two local minima around $u_1=0$ and
$u_1=1/2$ near the transition temperature.  Thus, during the transition,
the probability weight $e^{-f(\beta;u_1)}$ is described by two Gaussian
distributions around the minima.  This is a natural explanation of the
observed two-level nature.


The fact that the two-level nature of the transition in this model only
becomes apparent for rather large $N$ is not encouraging for numerical
studies of more complicated models with fermions.  However, the growth
of the specific heat with $N^2$ was apparent for smaller $N$, as seen in
Figure \ref{Fig:pols}, suggesting it is useful to monitor this
observable whenever possible if the nature of the transition is in
question. Such issues might be problematic in the class of
supersymmetric models especially the BMN model where a rich phase
structure is expected.

However, there is still room for discussion about the nature of the
observed transition.  As discussed in
\cite{Aharony:2003sx,Furuuchi:2003sy} by using state-counting analysis,
the Hagedorn behaviour is predicted in matrix models. In fact, we
observe that the $N$-dependence of the Polyakov loop at low-temperature
phase is well fit by the prediction based on the Hagedorn behaviour
\cite{Kovacik:2020cod}, where the finite-$N$ effects are large even for
very large $N$ in the transition region. Hence, there is little doubt
that the low-temperature phase of the model has fluctuations
characteristic of a Hagedorn transition.

In future work we plan to investigate the model for more general values
of $\mu$. Also we plan to return to the study of the D0--D4
Berkooz-Douglas model \cite{Filev:2015cmz,Asano:2016xsf,Asano:2016kxo}.
It would be natural to perform a similar study to the current one for
the bosonic version where one can investigate the effect of the
fundamental degrees of freedom on the system. This model becomes
especially interesting for $N_f=2N$ whose supersymmetric version is the
dimensional reduction to time of the superconformally invariant
four-dimensional model.  An initial study of this model was performed in
\cite{Asano:2018nol} and the exceptional behaviour of $N_f = 2 N$ was
noted.

\subsection*{Acknowledgment}
The authors wish to acknowledge the Irish Centre for High-End
Computing (ICHEC) for the provision of computational facilities and
support (Projects dsphy009c, dsphy010c and dsphy012c). The support from
Action MP1405 QSPACE of the COST foundation is gratefully
acknowledged.  Y.~Asano is supported by the JSPS Research Fellowship
for Young Scientists. S. Kov\'a\v{c}ik was supported by Irish Research
Council funding. The authors would like to thank G.~Bergner,
M.~Hanada, G.~Ishiki, T.~Morita and H.~Watanabe for valuable discussions.

\bibliographystyle{abbrv}

\begin{thebibliography}{99}

\bibitem{deWit:1988wri}
  B.~de Wit, J.~Hoppe and H.~Nicolai,
  ``On the Quantum Mechanics of Supermembranes,''
  Nucl.\ Phys.\ B {\bf 305} (1988) 545.
  doi:10.1016/0550-3213(88)90116-2

\bibitem{Banks:1996vh} 
  T.~Banks, W.~Fischler, S.~H.~Shenker and L.~Susskind,
  ``M theory as a matrix model: A Conjecture,''
  Phys.\ Rev.\ D {\bf 55}, 5112 (1997)
  [hep-th/9610043].

\bibitem{Berenstein:2002jq}
  D.~E.~Berenstein, J.~M.~Maldacena and H.~S.~Nastase,
  ``Strings in flat space and pp waves from N=4 super Yang-Mills,''
  JHEP {\bf 0204} (2002) 013
  doi:10.1088/1126-6708/2002/04/013
  [hep-th/0202021].

\bibitem{Kim:2006wg}
  N.~Kim and J.~H.~Park,
  ``Massive super Yang-Mills quantum mechanics: Classification and the relation to supermembrane,''
  Nucl.\ Phys.\ B {\bf 759} (2006) 249
  doi:10.1016/j.nuclphysb.2006.10.005
  [hep-th/0607005].

\bibitem{Aharony:2003sx}
  O.~Aharony, J.~Marsano, S.~Minwalla, K.~Papadodimas and M.~Van Raamsdonk,
  ``The Hagedorn - deconfinement phase transition in weakly coupled large N gauge theories,''
  Adv.\ Theor.\ Math.\ Phys.\  {\bf 8} (2004) 603
  doi:10.4310/ATMP.2004.v8.n4.a1
  [hep-th/0310285].

\bibitem{Furuuchi:2003sy}
  K.~Furuuchi, E.~Schreiber and G.~W.~Semenoff,
  ``Five-brane thermodynamics from the matrix model,''
  hep-th/0310286.

\bibitem{Semenoff:2005ei}
  G.~W.~Semenoff,
  ``Black holes and thermodynamic states of matrix models,''
  In *Shifman, M. (ed.) et al.: From fields to strings, vol. 3* 2009-2034

\bibitem{Aharony:2005ew}
  O.~Aharony, J.~Marsano, S.~Minwalla, K.~Papadodimas, M.~Van Raamsdonk and T.~Wiseman,
  ``The Phase structure of low dimensional large N gauge theories on Tori,''
  JHEP {\bf 0601} (2006) 140
  doi:10.1088/1126-6708/2006/01/140
  [hep-th/0508077].

\bibitem{Asano:2018nol}
  Y.~Asano, V.~G.~Filev, S.~Kov\'a\v{c}ik and D.~O'Connor,
  ``The non-perturbative phase diagram of the BMN matrix model,''
  JHEP {\bf 1807} (2018) 152
  doi:10.1007/JHEP07(2018)152
  [arXiv:1805.05314 [hep-th]].

\bibitem{Gregory:1993vy}
  R.~Gregory and R.~Laflamme,
  ``Black strings and p-branes are unstable,''
  Phys.\ Rev.\ Lett.\  {\bf 70} (1993) 2837
  doi:10.1103/PhysRevLett.70.2837
  [hep-th/9301052].

\bibitem{Gregory:1994bj}
  R.~Gregory and R.~Laflamme,
  ``The Instability of charged black strings and p-branes,''
  Nucl.\ Phys.\ B {\bf 428} (1994) 399
  doi:10.1016/0550-3213(94)90206-2
  [hep-th/9404071].

\bibitem{Aharony:2004ig}
  O.~Aharony, J.~Marsano, S.~Minwalla and T.~Wiseman,
  ``Black hole-black string phase transitions in thermal 1+1 dimensional
  supersymmetric Yang-Mills theory on a circle,''
  Class.\ Quant.\ Grav.\  {\bf 21} (2004) 5169
  doi:10.1088/0264-9381/21/22/010
  [hep-th/0406210].
  
\bibitem{Kawahara:2007fn}
  N.~Kawahara, J.~Nishimura and S.~Takeuchi,
  ``Phase structure of matrix quantum mechanics at finite temperature,''
  JHEP {\bf 0710} (2007) 097
  doi:10.1088/1126-6708/2007/10/097
  [arXiv:0706.3517 [hep-th]].
  
\bibitem{Filev:2015hia}
  V.~G.~Filev and D.~O'Connor,
  ``The BFSS model on the lattice,''
  JHEP {\bf 1605} (2016) 167
  doi:10.1007/JHEP05(2016)167
  [arXiv:1506.01366 [hep-th]].
  
\bibitem{Azuma:2014cfa}
  T.~Azuma, T.~Morita and S.~Takeuchi,
  ``Hagedorn Instability in Dimensionally Reduced Large-N Gauge Theories as Gregory-Laflamme and Rayleigh-Plateau Instabilities,''
  Phys.\ Rev.\ Lett.\  {\bf 113} (2014) 091603
  doi:10.1103/PhysRevLett.113.091603
  [arXiv:1403.7764 [hep-th]].

\bibitem{Mandal:2009vz}
  G.~Mandal, M.~Mahato and T.~Morita,
  ``Phases of one dimensional large N gauge theory in a 1/D expansion,''
  JHEP {\bf 1002} (2010) 034
  doi:10.1007/JHEP02(2010)034
  [arXiv:0910.4526 [hep-th]].

\bibitem{Bergner:2019rca}
  G.~Bergner, N.~Bodendorfer, M.~Hanada, E.~Rinaldi, A.~Sch\"afer and P.~Vranas,
  ``Thermal phase transition in Yang-Mills matrix model,''
  [arXiv:1909.04592 [hep-th]].  

\bibitem{Morita:2020liy}
  T.~Morita and H.~Yoshida,
  ``A Critical Dimension in One-dimensional Large-N Reduced Models,''
  arXiv:2001.02109 [hep-th].
  
\bibitem{Hanada:2018zxn}
  M.~Hanada, G.~Ishiki and H.~Watanabe,
  ``Partial Deconfinement,''
  JHEP {\bf 1903} (2019) 145
   Erratum: [JHEP {\bf 1910} (2019) 029]
  doi:10.1007/JHEP03(2019)145, 10.1007/JHEP10(2019)029
  [arXiv:1812.05494 [hep-th]].

\bibitem{Hanada:2016pwv}
  M.~Hanada and J.~Maltz,
  ``A proposal of the gauge theory description of the small Schwarzschild black hole in AdS$_5\times$S$^5$,''
  JHEP {\bf 1702} (2017) 012
  doi:10.1007/JHEP02(2017)012
  [arXiv:1608.03276 [hep-th]].
  
\bibitem{Berenstein:2018lrm}
  D.~Berenstein,
  ``Submatrix deconfinement and small black holes in AdS,''
  JHEP {\bf 1809} (2018) 054
  doi:10.1007/JHEP09(2018)054
  [arXiv:1806.05729 [hep-th]].

\bibitem{Hanada:2019czd}
   M.~Hanada, A.~Jevicki, C.~Peng and N.~Wintergerst,
   ``Anatomy of Deconfinement,''
   arXiv:1909.09118 [hep-th].

\bibitem{Emparan:2018bmi}
  R.~Emparan, R.~Luna, M.~Mart\'inez, R.~Suzuki and K.~Tanabe,
  ``Phases and Stability of Non-Uniform Black Strings,''
  JHEP {\bf 1805} (2018) 104
  doi:10.1007/JHEP05(2018)104
  [arXiv:1802.08191 [hep-th]].

\bibitem{Dias:2017uyv}
  \'O.~J.~C.~Dias, J.~E.~Santos and B.~Way,
  ``Localised and nonuniform thermal states of super-Yang-Mills on a circle,''
  JHEP {\bf 1706} (2017) 029
  doi:10.1007/JHEP06(2017)029
  [arXiv:1702.07718 [hep-th]].

\bibitem{Cardona:2018shd}
  B.~Cardona and P.~Figueras,
  ``Critical Kaluza-Klein black holes and black strings in D = 10,''
  JHEP {\bf 1811} (2018) 120
  doi:10.1007/JHEP11(2018)120
  [arXiv:1806.11129 [hep-th]].

\bibitem{Ammon:2018sin}
  M.~Ammon, M.~Kalisch and S.~Moeckel,
  ``Notes on ten-dimensional localized black holes and deconfined states in two-dimensional SYM,''
  JHEP {\bf 1811} (2018) 090
  doi:10.1007/JHEP11(2018)090
  [arXiv:1806.11174 [hep-th]].

\bibitem{Witten:1998zw}
  E.~Witten,
  ``Anti-de Sitter space, thermal phase transition, and
  confinement in gauge theories,''
  Adv.\ Theor.\ Math.\ Phys.\  {\bf 2} (1998) 505
  doi:10.4310/ATMP.1998.v2.n3.a3
  [hep-th/9803131].

\bibitem{Asano:2019pre}
    Y.~Asano and D.~O'Connor
\textit{In preparation.}


\bibitem{Hadizadeh:2004bf}
  S.~Hadizadeh, B.~Ramadanovic, G.~W.~Semenoff and D.~Young,
  ``Free energy and phase transition of the matrix model on a plane-wave,''
  Phys.\ Rev.\ D {\bf 71} (2005) 065016
  doi:10.1103/PhysRevD.71.065016
  [hep-th/0409318].

\bibitem{Takeuchi:2017wii}
  S.~Takeuchi,
  ``D-dependence of the gap between the critical temperatures in
  the one-dimensional gauge theories,''
  Eur.\ Phys.\ J.\ C {\bf 79} (2019) no.7,  548
  doi:10.1140/epjc/s10052-019-6941-y
  [arXiv:1712.09261 [hep-th]].
    
\bibitem{DelgadilloBlando:2007vx}
  R.~Delgadillo-Blando, D.~O'Connor and B.~Ydri,
  ``Geometry in Transition: A Model of Emergent Geometry,''
  Phys.\ Rev.\ Lett.\  {\bf 100} (2008) 201601
  doi:10.1103/PhysRevLett.100.201601
  [arXiv:0712.3011 [hep-th]].
  
\bibitem{DelgadilloBlando:2008vi}
  R.~Delgadillo-Blando, D.~O'Connor and B.~Ydri,
  ``Matrix Models, Gauge Theory and Emergent Geometry,''
  JHEP {\bf 0905} (2009) 049
  doi:10.1088/1126-6708/2009/05/049
  [arXiv:0806.0558 [hep-th]].

\bibitem{Costa:2014wya}
  M.~S.~Costa, L.~Greenspan, J.~Penedones and J.~Santos,
  ``Thermodynamics of the BMN matrix model at strong coupling,''
  JHEP {\bf 1503} (2015) 069
  doi:10.1007/JHEP03(2015)069
  [arXiv:1411.5541 [hep-th]].
  
\bibitem{DelgadilloBlando:2012xg}
  R.~Delgadillo-Blando and D.~O'Connor,
  ``Matrix geometries and Matrix Models,''
  JHEP {\bf 1211} (2012) 057
  doi:10.1007/JHEP11(2012)057
  [arXiv:1203.6901 [hep-th]].

\bibitem{Fisher:1982xt} 
  M.~E.~Fisher and A.~N.~Berker,
  ``Scaling for first-order transitions in thermodynamic and finite systems,''
  Phys.\ Rev.\ B {\bf 26}, 2507 (1982).
  doi:10.1103/PhysRevB.26.2507

\bibitem{Landau:1984}
  D.~P.~Landau and K.~Binder,
  ``Finite-size scaling at first-order phase transitions''
  Phys.\ Rev.\ B {\bf 30} (1984) 1477.
  doi:10.1103/PhysRevB.30.1477
  
\bibitem{Challa:1986sk}
  M.~S.~S.~Challa, D.~P.~Landau and K.~Binder,
  ``Finite size effects at temperature driven first order transitions,''
  Phys.\ Rev.\ B {\bf 34} (1986) 1841.
  doi:10.1103/PhysRevB.34.1841
  
\bibitem{Glaser:2017sbe}
  L.~Glaser, D.~O'Connor and S.~Surya,
  ``Finite Size Scaling in 2d Causal Set Quantum Gravity,''
  Class.\ Quant.\ Grav.\  {\bf 35} (2018) no.4,  045006
  doi:10.1088/1361-6382/aa9540
  [arXiv:1706.06432 [gr-qc]].

\bibitem{Filev:2015cmz}
  V.~G.~Filev and D.~O'Connor,
  ``A Computer Test of Holographic Flavour Dynamics,''
  JHEP {\bf 1605} (2016) 122
  doi:10.1007/JHEP05(2016)122
  [arXiv:1512.02536 [hep-th]].
  
\bibitem{Asano:2016xsf}
  Y.~Asano, V.~G.~Filev, S.~Kov\'a\v{c}ik and D.~O'Connor,
  ``The Flavoured BFSS Model at High Temperature,''
  JHEP {\bf 1701} (2017) 113
  doi:10.1007/JHEP01(2017)113
   [arXiv:1605.05597 [hep-th]].

\bibitem{Asano:2016kxo}
  Y.~Asano, V.~G.~Filev, S.~Kov\'a\v{c}ik and D.~O'Connor,
  ``A computer test of holographic favour dynamics. Part II,''
  JHEP {\bf 1803} (2018) 055
  doi:10.1007/JHEP03(2018)055
  [arXiv:1612.09281 [hep-th]].

\bibitem{Kovacik:2020cod}
  S.~Kov\'a\v{c}ik, D.~O'Connor and Y.~Asano,
  ``The nonperturbative phase diagram of the bosonic BMN matrix model,''
  arXiv:2004.05820 [hep-th].

  
\end{thebibliography}

\end{document}